\documentclass[a4paper, 11pt]{article}    % Specifies the document style.
\usepackage{latexsym}
\usepackage{amssymb}
\usepackage{amsmath}
\usepackage{amsfonts}
\usepackage{graphicx}
\usepackage{float}
\usepackage[english]{babel}

\newcommand{\quota}[1]{``#1''}

\title{ \bf Funded Bilateral Valuation Adjustment}  % Declares the document's title.
\author{ Lorenzo Giada \\ Banco Popolare, Verona \\ {\tt lorenzo.giada@gmail.com}$^*$ \and Claudio Nordio \\ Banco Popolare, Verona \\ {\tt c.nordio@gmail.com}\footnote{This paper reflects the authors' opinions and
not necessarily those of their employers.}}

\begin{document}           % End of preamble and beginning of text.
\maketitle                 % Produces the title.

\begin{center}
\tt \Large Working paper
\end{center}

\vspace{10mm}

\begin{abstract}
\it \noindent
We show how the cost of funding the collateral in a particular set up can be equal to the Bilateral Valuation Adjustment with the \quota{funded} probability of default, leading to the definition of a Funded Bilateral Valuation Adjustment (FBVA). That set up can also be viewed by an investor as an effective way to restructure the counterparty risk arising from an uncollateralized transaction with a counterparty, mitigating or even avoiding entirely the additional capital charge introduced by the new Basel III framework.
\end{abstract}

\bigskip

{\bf JEL} Classification codes: G12, G13 

{\bf AMS} Classification codes: 62H20, 91B70 

\bigskip

{\bf Keywords:} Counterparty risk, Credit Valuation Adjustment, Funding Valuation Adjustment, Bilateral Valuation Adjustment, Debit Valuation Adjustment, CVA, FVA, DVA, Basel III, Restructuring Counterparty risk, ISDA, CSA, One way CSA.

\vspace{15mm}

\section{Summary}
It is well known \cite{brigo2} that the mark-to-market of derivatives between counterparties that do not post collateral includes terms that take into account the possibility that some counterparty defaults before the end of the contract. These quantities represent the so-called Bilateral Credit Valuation Adjustment (BCVA) and Debit Valuation Adjustment (BDVA), and must include the effect of the first-to-default.

On the other hand, if a collateral agreement is in place between the counterparties, neglecting the residual gap risk or assuming perfect collateralization as discussed below, it is accepted that the mark-to-market of the derivative is simply the present value of the contractual cash flows. In the following we show a way to structure a collateralized transaction such that, if one takes into account the costs of funding the collateral, a component representing counterparty risk can be recovered, that is equivalent in its form to the uncollateralized BCVA$-$BDVA, but with the default probabilities now obtained from the Asset Swap spreads of the parties rather than from their CDS spreads. We call this term the Funded Bilateral Valuation Adjustment (FBVA).

Moreover, the collateralized transaction that originates the FBVA can also be viewed as an effective way for an investor to restructure the counterparty risk of an uncollateralized transaction, reducing or avoiding the additional capital charge defined by the Basel III framework. Therefore the set up we introduce below might be criticized as a way to bypass Basel III, a point we will discuss below. As far as we know, however, a financial institution is legitimate to adopt it and then the issue will have to be addressed by the regulators.

\paragraph{This paper and the previous literature} Our framework is introduced in Section \ref{sNot} where we follow the notation of e.g.~\cite{brigo2}; in particular we will adopt the risk-free close-out amount, without affecting the validity of our results. We leave to a future work the extention to more general close-out conventions following \cite{brigo4}. In Section \ref{sUni}, while we show how our main result works in the unilateral case, we highlight that the present value of the cashflows related to margining the collateral is closely connected to the CVA, a result somewhat obvious but left implicit by the literature, see for instance \cite{albanese} where margin lending are priced, and refer to \cite{brigo1} for an extensive treatment of the funding valuation problem and a complete survey of the literature on this topic. Unlike the latter, we will not address the subtleties of the whole inclusion of the cost of funding while pricing a transaction, being interested in the less extensive scope of including the cost of collateral in a new structuring. The spirit of our approach is similar to \cite{prampolini} where the authors, however, examine the cost of funding issue in fundamental but simple transactions with upfront, i.e. synthetic loans. Our main result in the general, bilateral structure is illustrated in \ref{sBil}, and all the calculations, though rather simple, are postponed to the Appendix. The implication of our results are discussed in Section \ref{sDisc}.

\section{Notation}\label{sNot}
The value for $B$ of a derivative contract struck between two defaultable counterparties $A$ and $B$, when a default risk-free close-out amount convention is adopted, is
\begin{equation}\label{fairvalue1}
V^{AB}_B(t_0) = V^0_B(t_0) -BCVA_B(t_0,T)+BDVA_B(t_0,T),
\end{equation}
%\begin{eqnarray}
%V^{AB}_B(t_0) &=& V^0_B(t_0) - \mathbb{E}\left\{\left. L_A\mathbb{I}_A(t_0,T) D(t_0,\tau_A) \left[ V_B^0(\tau_A) \right]^+
%\right| t_0 \right\} \nonumber\\
%& & + \mathbb{E}\left\{\left. L_B\mathbb{I}_B(t_0,T) D(t_0,\tau_B) \left[ -V_B^0(\tau_B) \right]^+ \right| t_0 \right\},\label{fairvalue}
%\end{eqnarray}
where
\begin{equation}\label{BCVA}
BCVA_B(t,T) = \mathbb{E}\left\{\left. L_A\mathbb{I}_A(t,T) D(t,\tau_A) \left[ V_B^0(\tau_A) \right]^+ \right| t \right\},
\end{equation}
\begin{equation}\label{BDVA}
BDVA_B(t,T) = \mathbb{E}\left\{\left. L_B\mathbb{I}_B(t,T) D(t,\tau_B) \left[ -V_B^0(\tau_B) \right]^+ \right| t \right\},
\end{equation}
and $D(t_1,t_2)$ is the stochastic discount factor between two dates, $\tau_X$ is the default time of counterparty $X$, $\mathbb{I}_A(t_1,t_2)=\mathbb{I}_{t_1<\tau_A<\min(\tau_B,t_2)}$, $\mathbb{I}_B(t_1,t_2)=\mathbb{I}_{t_1<\tau_B<\min(\tau_A,t_2)}$, $T$ is the last payment date of the derivative, $L_X$ the loss given default of counterparty $X$, and $V^0_X(t)$ is the value of the equivalent default risk free derivative as seen from $X$, e.g.
\begin{equation}
V_B^0(t) = \mathbb{E}\left\{\left. \Pi_B(t,T) \right| t \right\} =
\mathbb{E}\left\{\left. \sum_{i=1}^N D(t,T_i)C_i(T_i) \right| t \right\},
\end{equation}
with $T_1>t$, $T_N=T$, and $C_i$ the cashflow paid in $T_i$ that depends on the values of the risk factors (e.g. interest rates, stock prices) already known in $T_i$. We also define $\tau=\min(\tau_A,\tau_B)$, and we will use the notation $\mathbb{P}_A(t_1,t_2)=\mathbb{E}\left\{\mathbb{I}_A(t_1,t_2)\right\}$, and analogous for $B$, for the unconditioned probabilities. We will omit conditioning on $t_0$ henceforth.

%The last two terms in eq.~(\ref{fairvalue}) define the BCVA and BDVA respectively as seen from $B$,
%so that we can write the well known formula

A discrete equivalent of each of the two formulas (\ref{BCVA}) and (\ref{BDVA}) can be obtained assuming that the default can happen (or be observed) only at specific dates $\{ t_i\}_{i=1,...,M}$ with $t_0=t$ and $t_M=T$ (e.g.~the dates in which cash flows are exchanged):

\begin{equation}\label{dBCVA}
BCVA_B(t,T) = \sum_{k=1}^M  
\mathbb{E}\left\{\left.  L_A \mathbb{I}_{t_{k-1}\le\tau_A\le\min\left(t_k,\tau_B\right)} D(t_0,t_{k-1}) \left[ V_B^0(t_{k-1}) \right]^+ \right| t \right\},
\end{equation}
\begin{equation}\label{dBDVA}
BDVA_B(t,T) = \sum_{k=1}^M  
\mathbb{E}\left\{\left.  L_B \mathbb{I}_{t_{k-1}\le\tau_B\le\min\left(t_k,\tau_A\right)} D(t_0,t_{k-1}) \left[ V_A^0(t_{k-1}) \right]^+ \right| t \right\}.
\end{equation}

\section{Funded Bilateral Valuation Adjustment (FBVA)}\label{sFBVA}
In order to enter into an uncollateralized transaction with an initial default risk free mark-to-market equal to zero, a counterparty has to receive its Bilateral Valuation Adjustment BVA = BCVA$-$BDVA, since the fair value of the derivative is $-$BVA. On the other hand, if the deal is collateralized, there is apparently no need of such a payment. We show in the following that by taking into account the flows related to the borrowing and lending of the collateral in a particular set up, we recover payments that are equivalent, up to the definition of default probability, to the BVA.

In what follows we assume perfect collateralization, that is, both minimum transfer amount and threshold are set to zero in the standard CSA agreement, and the amount of the collateral is computed daily. We neglect the gap risk of intraday price changes of the derivative and allow rehypotecation of collateral.

The next paragraph is devoted to illustrate the core idea in a simplified setting, where one party is risk-free and then has to receive the CVA to enter into an uncollateralized transaction with a defaultable counterparty. In a subsequent section we obtain the general formulas for two defaultable counterparties.

\subsection{The unilateral case}\label{sUni}
Assume for the moment that one of the counterparties $-B-$ is riskless, hence does not need to post collateral, while the other $-A-$ posts cash collateral. We would like to take into account the costs and profits related to this collateral. At any particular time $t$, A has to borrow an amount of cash  $\left[ V_B^0(t) \right]^+$ equal to the mark to market of the derivative, if positive for $B$. $A$ has to pay its cost of funding on this amount, that can be quantified in a risk free rate (hereafter the Eonia rate $\varepsilon$) plus a constant funding spread $F_A$ which can be thought of as being related to the asset swap spread of $A$. This flow of payments lasts as long as $A$ does not default. Then $A$ posts the collateral to $B$, usually receiving the risk free rate. Also these payments are interrupted at the default of $A$. The counterparty $B$ is now long the collateral, and can invest it either on the market, or with $A$, providing precisely the amount of cash that she needed in the first place to provide the collateral. In this case $B$ receives $\varepsilon + F_A$.
	
\begin{figure}[h]
\centering
\includegraphics[width=0.50\textwidth]{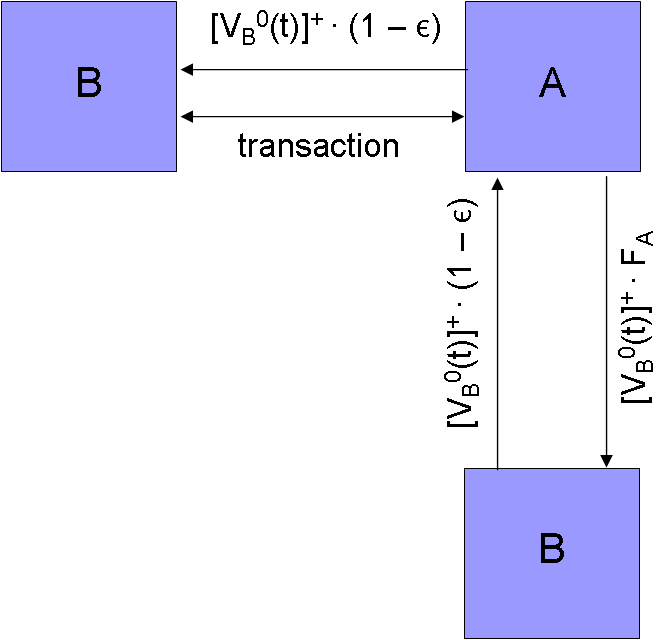}
\caption{Unilateral set up, with the cashflows in case of survivalship of $A$. $B$ is risk free and $A$ posts collateral receiving Eonia on it (upper horizontal leg); $A$ borrows the collateral from $B$ and pays Eonia plus its own funding spread (vertical legs).\label{fig1}}
\end{figure}

Figure \ref{fig1} represents schematically this last set up, with the cashflows in case of survivalship of $A$. The horizontal legs refer to the collateralized transaction, the vertical to the loan. The {\it transaction} leg represents the derivative transaction, including the exchange of $V_B^0(t_0)$, whereas $\left[ V_B^0(t) \right]^+\cdot\left(1-\epsilon\right)$ is a condensed notation for the margining cashflows (at any $t$, $A$ has posted the amount of cash $\left[ V_B^0(t) \right]^+$ and receives Eonia interests on it). The same margining cashflows are part of the loan, and we will refer to both as the Margin Cashflows\footnote{including the cashflows in case of default, see the Appendix for the details} hereafter. The leg labeled with $\left[ V_B^0(t) \right]^+\cdot F_A$ represents the cost of funding for $A$ over the risk-free rate.

Since all risk free flows cancel, assuming the periodic payments of the funding spread by $A$ are in advance, the total revenue for $B$ to set up this transaction is
\begin{equation}\label{CVAcap}
F_A \sum_{k=1}^M \theta_k
\mathbb{E}\left\{\mathbb{I}_{\tau_A>t_{k-1}} D(t_0,t_{k-1}) \left[ V_B^0(t_{k-1}) \right]^+ \right\}
\end{equation}
where $\theta_k = t_k-t_{k-1}$ and the dates $t_k$ are now the margining dates in which collateral can be requested.

On the other hand, from the point of view of $B$ the quantity (\ref{CVAcap}) is a receive leg whose present value must balance the expected loss of the loan, e.g. the present value of the related Margin Cashflows as seen from $B$. We call such expected loss the Funded Credit Valuation Adjustment (FCVA). As the Margin Cashflows under the assumptions of no gap risk cancel the counterparty risk of the original uncollateralized transaction, the no-arbitrage principle leads to conclude that (\ref{CVAcap}) is worth the CVA of the derivative. However, the functional form of FCVA is equivalent (up to the estimation of the default probabilities, see the Section \ref{sDisc} below) to the unilateral CVA computed for the uncollateralized transactions, see Appendix \ref{appB} for the derivation: 
\begin{equation}\label{FCVA}
FCVA = \sum_{k=1}^M \mathbb{E}\left\{L_A \mathbb{I}_{t_{k-1}< \tau_A \le t_k} D(t_0,t_{k-1}) \left[V^0_B(t_{k-1})\right]^+\right\},
\end{equation}
and comparing it with (\ref{CVAcap}) one may obtain the fair value of $F_A$. Let us point out that FCVA is part of the \emph{fair value} of the loan subscribed by $A$ to fund the collateral, and is not a component of the price of the derivative. It is a cost to $A$ related to the derivative in the sense that, had she not entered into the derivative transaction, she would not have requested the loan. The collateral mechanism has moved the CVA from the original transaction to a loan. Hereafter, we will refer to such loan as a Margin Loan.

\subsection{The bilateral case}\label{sBil}
The next step is to assume that also $B$ is defaultable, so in the Margin Loan above we have to take into account that the possible default of $B$ implies the early termination of the loan. We obtain the following estimate for the cost of the loan over the Eonia rate, see the Appendix for the derivation:
\begin{equation}\label{FBCVA}
FBCVA=\sum_{k=1}^M \mathbb{E}\left\{L_A \mathbb{I}_{t_{k-1}< \tau_A \le \min\left(t_k,\tau_B\right)} D(t_0,t_{k-1}) \left[V^0_B(t_{k-1})\right]^+\right\},
\end{equation}
which has to be worth as
\begin{equation}
F_A \sum_{k=1}^M \theta_k
\mathbb{E}\left\{\mathbb{I}_{\tau_A>t_{k-1}}\mathbb{I}_{\tau_B>t_{k-1}} D(t_0,t_{k-1}) \left[ V_B^0(t_{k-1}) \right]^+ \right\}
\end{equation}
to make the loan fair, obtaining a spread $F_A$ different from that of the unilateral case.

\begin{figure}[t]
\centering
\includegraphics[width=0.45\textwidth]{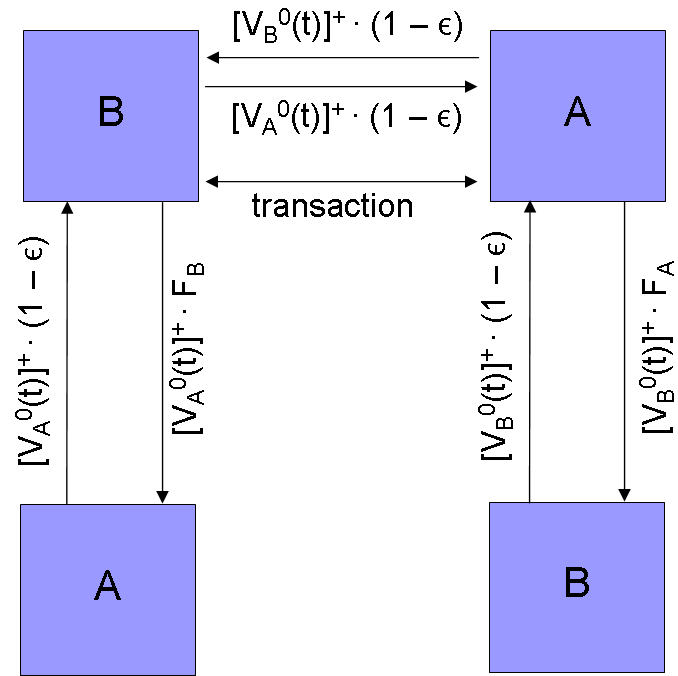}
\caption{Bilateral set up. Both $A$ and $B$ post collateral that is reinvested with the same counterparty.\label{fig2}}
\end{figure}

\begin{figure}[t]
\centering
\includegraphics[width=0.85\textwidth]{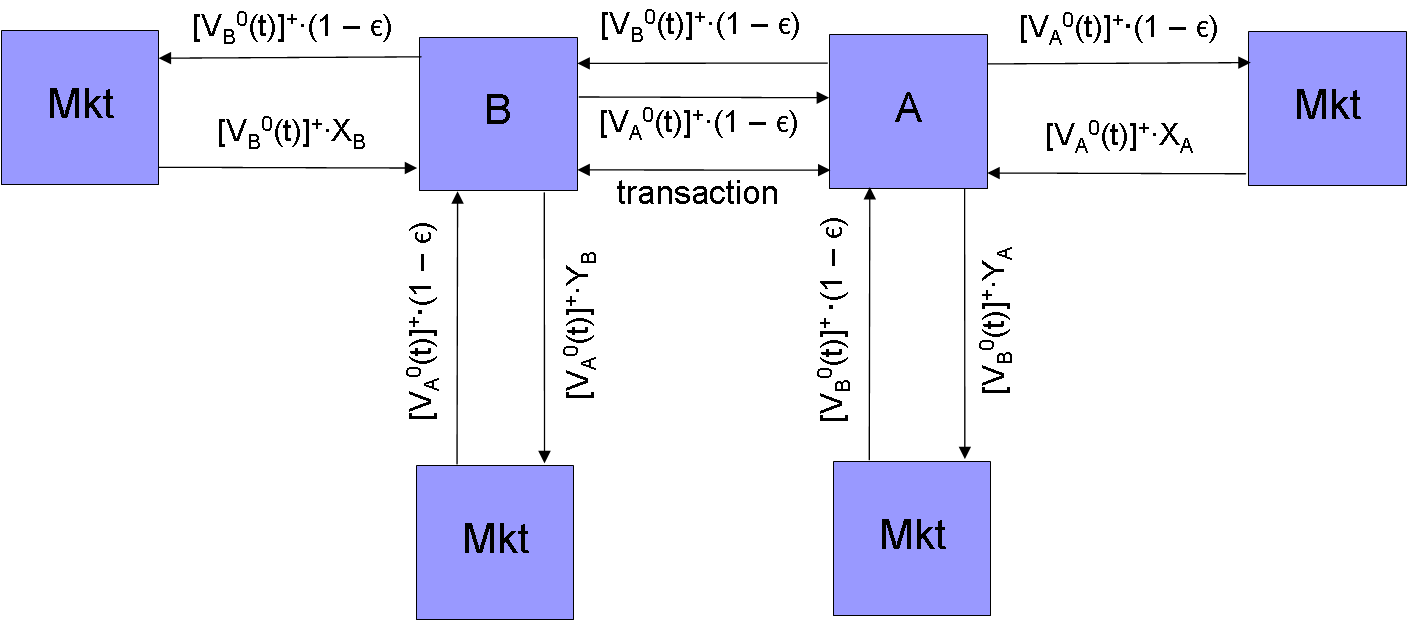}
\caption{Bilateral set up. Both $A$ and $B$ post collateral, that is reinvested in the market. The funding spreads $Y_A$ and $Y_B$ are related to their respective funding spreads, the investment spreads $X_A$ and $X_B$ depend from the credit risk of the borrowers\label{fig3}}
\end{figure}

Moreover, $B$ has now to post collateral to $A$. In order to do this, $B$ has to borrow an amount of cash equal to $\left[ V_A^0(t) \right]^+$, and has to pay on it the risk-free rate plus its funding spread. On the other hand, $A$ can reinvest the collateral she has received. In figure \ref{fig3} it is represented a general situation where $A$ and $B$ borrow on the market at spread $Y_A$ or $Y_B$ related to their funding spread, and invest on the market at an a priori unknown spread $X_A$ or $X_B$ related to the credit worthiness of the respective borrowers. Again, we will refer instead to the case illustrated in figure \ref{fig2} whith $A$ and $B$ funding each other's collateral. Therefore, in analogy to the above, we obtain the total cost for $B$ related to borrowing the collateral from $A$ as
\begin{equation}\label{FBCVA}
FBDVA=\sum_{k=1}^M \mathbb{E}\left\{L_B \mathbb{I}_{t_{k-1}< \tau_B \le \min\left(t_k,\tau_A\right)} D(t_0,t_{k-1}) \left[V^0_A(t_{k-1})\right]^+\right\}.
\end{equation}
which has to be worth as
\begin{equation}
F_B \sum_{k=1}^M \theta_k
\mathbb{E}\left\{\mathbb{I}_{\tau_A>t_{k-1}}\mathbb{I}_{\tau_B>t_{k-1}} D(t_0,t_{k-1}) \left[ V_A^0(t_{k-1}) \right]^+ \right\}
\end{equation}
to obtain the fair spread $F_B$, and our main result follows by concluding that $B$ either receives BVA to enter into the original, uncollateralized transaction, or receives - through the funding spread cashflows - an economic equivalent of FBVA = FBCVA$-$FBDVA to collateralize the transaction through the two Margin Loans.

\section{Discussion}\label{sDisc}

The sets of transactions in figure \ref{fig1} or figure \ref{fig2} have interesting features and implications discussed in the following.

\paragraph{Funded vs.~unfunded transactions} We have shown above that the FBVA has the same functional form of BVA, but we argue that the evaluation of the former should involve the default probabilities implied by the Asset Swap spread rather than by the CDS spread curve, because the FBVA is related to unsecured loans. This is more evident in the unilateral case if one considers a synthetic loan derivative, with a bullet cashflow at maturity, where the Margin Loan is reduced to a simple loan with that cashflow as notional. Therefore might be valuable for an investor to enter into the collateralized transaction assisted by the two Margin Loans instead of the original uncollateralized transaction, because the latter has the same cashflows of the former, but the components of FBVA, namely FCVA and FDVA, are generally smaller than the unfunded BVA (this follows from the generally positive basis between CDS and Asset Swap spread).

\paragraph{No new liquidity needed} Remarkably, if rehypotecation of the collateral is allowed, the Margin Loans represent the only way to fund the transaction without the need of injecting new liquidity in the market, a problem that has led practitioners to manage funding strategies, see for instance \cite{albanese}.

\paragraph{Mitigating Basel III capital charges} Following the above, restructuring an uncollateralized transaction in a set of transactions including margining and the Margin Loans is a valuable way to mitigate the counterparty risk arising from the original transaction, the cashflows being exactly the same in case of either survivalship or default of one party (up to the running payment of the spread, but this difference could be remedied by considering a running paid BVA in the original transaction). Indeed the components of FBVA, namely FCVA and FDVA, are generally not as risky as the unfunded BVA, because the Asset Swap spread used in the former is generally less risky than the CDS spread used in the latter (this is due to the riskiness of the basis between the two, generally positively correlated with the two).

\paragraph{Avoiding Basel III capital charges and criticism} Furthermore, if one party designates the Margin Loan at its historical cost instead of its fair value, she avoids entirely the  capital charges determined by the Basel III framework. This is a consequence of having transformed the counterparty risk into a classical credit risk arising from the Margin Loans. The Basel II capital charge of the latter is essentially the same of the former, as they have the same exposure at default (equal to the mark-to-market, if positive, at the time of default), whereas the additional requirements are not applicable being uniquely related to bilateral transactions and their impact on the balance sheet. The structure shown in figure \ref{fig2} does not tranfer the credit risk and then could be viewed simply as a cunning way to bypass Basel III, but it is only a straightforward consequence of the accounting rules in force today, which imply different impact to the balancesheet for different but financially equivalent instruments. However, regulators should consider that a financial institution could frustrate the effort of Basel Committee toward capital resilience and counterparty risk hedging.

\paragraph{The case of one-way collateral} The case of a risk free counterparty in figure \ref{fig1} could be thought of the restructuring of a \quota{one-way} collateral assisted transaction (so called \quota{one-way} CSA in the ISDA framework), where the party $B$ already posts collateral to $A$. From the point of view of $B$, this fully counterbalances the (bilateral) DVA but leaves the (bilateral) CVA not null.

\paragraph{Meaning of the own risk (also for a lender)} The transactions \ref{fig1} or \ref{fig2} suggest a possible interpretation of the DVA viewed by a party as a component of the cost of funding the collateral owned by that party and posted to the other one. As known, the own risk impacts the bilateral CVA through the first-to-default effect; in the funded bilateral CVA, the first-to-default effect is a consequence of the early termination of the loan triggered by one's own default, which implies that the riskier a lender is, the lower spread it will request to compensate the expected loss of the borrower until its own default.\newline

So far we have not considered the residual gap risk which in practice arises for each new margined transaction. It is usual to hedge this risk by posting to the counterparty an Independent Amount (i.e.~an amount that does not depend to the mark-to-market of the transaction). Alternatively, it could be convenient to mitigate it through netting with the Margin Loans themselves\footnote{we are grateful to an undisclosed counterparty for having highlighted this issue}. This could be realized in practice by introducing a \quota{cross-product netting} clause between the collateralized transaction and the Margin Loans. However this clause increases the credit risk related to the loans.\newline

We conclude observing that the above \quota{regulamentary arbitrage} leads to a clash of two Titans, respectively the fair value accounting principle of a derivative and the possibility to show a loan at its historical cost. As it is not easy to give up either of the two principles, it could be appropriate to distinguish (at least in the capital requirements regulation) the purpose for which the parties has entered into a derivative transaction. On one hand a new set of rules could impose the fair value accounting of the Margin Loans if associated to a derivative transaction negotiated for trading purposes; on the other hand it could allow its accounting at historical cost when related to real hedging/held to maturity transactions.

\section*{Acknowledgments}
We are grateful to Damiano Brigo for helpful discussions and Carlo Palego for encouraging our research.

\appendix

\section{Pricing the Margin Cashflows}\label{appB}
In what follows we make use of the simplified notation $V^+_X(t)=\left[V^0_X(t)\right]^+$.
\paragraph{Unilateral case: B default-risk free} The Margin Cashflows of the Margin Loan valued in $t_0$ from the point of view of $B$ are given by the stochastic quantity
\begin{eqnarray}
&&-V^+_B(t_0)\mathbb{I}_{\tau_A>t_0}+D(t_0,t_M) V^+_B(t_M)\mathbb{I}_{\tau_A>t_M}- \nonumber \\ && \sum_{k=1}^M\left[V^+_B(t_k)- V^+_B(t_{k-1}) \right] D(t_0, t_k) \mathbb{I}_{\tau_A>t_k} + \sum_{k=1}^M V^+_B(t_{k-1})D(t_0, t_k)\theta_k\varepsilon_{k-1}\mathbb{I}_{\tau_A>t_k}
+ \nonumber\\ && R_A\sum_{k=1}^M  V^+_B(t_{k-1}) (1+\theta_k\varepsilon_{k-1})\mathbb{I}_{t_{k-1}< \tau_A \le t_k}D(t_0,t_k),
\end{eqnarray}
where $\varepsilon_{k-1}$ is the eonia rate (or any other risk-free rate) for the period $(t_{k-1},t_k)$, and $t_M=T$ is the maturity of the transaction. The first four elements of the above sum represent the cashflows in case of survivalship of $A$; the first three come from the margination, the fourth is the interest rate leg. The fifth term is the recovery in case of default of $A$ and includes the recovery of the interest rate matured in the relevant period. Observe first that
\begin{equation}
D(t_0,t_k)(1+\theta_k\varepsilon_{k-1}) = D(t_0,t_{k-1}),
\end{equation}
and then that
\begin{equation}
\mathbb{I}_{\tau_A>t_{k-1}}-\mathbb{I}_{\tau_A>t_k}
=\mathbb{I}_{t_{k-1}< \tau_A \le t_k}.
\end{equation}
We arrive to the expression
\begin{equation}
-\sum_{k=1}^M V^+_B(t_{k-1}) D(t_0,t_{k-1})\left(1-R_A\right) \mathbb{I}_{t_{k-1}< \tau_A \le t_k},
\end{equation}
and taking the expected value we obtain the following expected loss
\begin{equation}
\sum_{k=1}^M \mathbb{E}\left\{L_A \mathbb{I}_{t_{k-1}< \tau_A \le t_k} D(t_0,t_{k-1}) V^+_B(t_{k-1})\right\}.
\end{equation}
i.e.~the BCVA discretized as in (\ref{dBCVA}) and reduced to the unilateral case, evaluated at $t_0$.

\paragraph{Bilateral case: B defaultable} The algebra is similar to the above, but the first to default effect has to be considered. The Margin Cashflows are now given by the stochastic quantity
\begin{eqnarray}
&&-V^+_B(t_0)\mathbb{I}_{\tau_A>t_0}\mathbb{I}_{\tau_B>t_0}+D(t_0,t_M) V^+_B(t_M)\mathbb{I}_{\tau_A>t_M}\mathbb{I}_{\tau_B>t_M}+
\nonumber \\ && 
\sum_{k=1}^M V^+_B(t_{k-1})D(t_0, t_k)\theta_k\varepsilon_{k-1}\mathbb{I}_{\tau_A>t_k}\mathbb{I}_{\tau_B>t_k}- \nonumber \\ &&  
\sum_{k=1}^M\left[V^+_B(t_k)- V^+_B(t_{k-1}) \right] D(t_0, t_k) \mathbb{I}_{\tau_A>t_k}\mathbb{I}_{\tau_B>t_k} + \nonumber \\ && 
R_A\sum_{k=1}^M  V^+_B(t_{k-1}) (1+\theta_k\varepsilon_{k-1})D(t_0,t_k)\mathbb{I}_{t_{k-1}< \tau_A \le \min\left(t_k,\tau_B \right)}
+ \nonumber \\ &&
\sum_{k=1}^M  V^+_B(t_{k-1}) (1+\theta_k\varepsilon_{k-1})D(t_0,t_k)\mathbb{I}_{t_{k-1}< \tau_B \le \min\left(t_k,\tau_A \right)}.
\end{eqnarray}
The terms in the above sum come from the contribution of
\begin{itemize}
\item{the survivalship of both $A$ and $B$ in the $k$-th period: $\tau_A>t_{k}$ and $\tau_B>t_{k}$};
\item{the default of $A$ first in the $k$-th period: $t_{k-1} \le \tau_A \le \min\left(t_k,\tau_B\right)$}
\item{the default of $B$ first in the $k$-th period: $t_{k-1} \le \tau_B  \le \min\left(t_k,\tau_A\right)$}
\end{itemize}
the survival case. A little algebra, using also the helpful equation
\begin{equation}\nonumber
\mathbb{I}_{t_{k-1}< \tau_A \le t_k}\cdot\mathbb{I}_{t_{k-1}< \tau_B \le t_k}=\mathbb{I}_{t_{k-1}< \tau_A < \tau_B\le t_k}+\mathbb{I}_{t_{k-1}< \tau_B < \tau_A\le t_k}
\end{equation}
leads to the following expression for the expected loss
\begin{equation}
\sum_{k=1}^M \mathbb{E}\left\{L_A \mathbb{I}_{t_{k-1}< \tau_A \le \min\left(t_k,\tau_B\right)} D(t_0,t_{k-1}) V^+_B(t_{k-1})\right\},
\end{equation}
i.e.~the BCVA discretized as in (\ref{dBCVA}) and evaluated at $t_0$.

\end{document}